\documentclass[a4paper]{jpconf}
\usepackage{amssymb}
\usepackage{graphicx}
\usepackage[numbers,square,sort&compress]{natbib}
\usepackage[section]{placeins}
\usepackage{booktabs, array}
\usepackage{multirow}
\usepackage{hepnicenames}
\usepackage{tikz}
\usetikzlibrary{arrows,positioning}
\usepackage{wrapfig}
\usepackage{hyperref}
\usepackage[noabbrev]{cleveref}

\usepackage[
  locale=US,
  separate-uncertainty=true,
  per-mode=symbol-or-fraction,
  group-minimum-digits={3},
]{siunitx}
\newcommand{\DKpipipi}{\mbox{\(\PDzero \rightarrow \PKp\Ppiminus\Ppiplus\Ppiminus\ \)}}

\usepackage{lineno}

\begin{document}

\title{Amplitude analysis of four-body decays using a massively-parallel fitting framework}

\author{C Hasse$^{1,2,3}$, J Albrecht$^2$, A A Alves Jr.$^3$, P d'Argent$^4$, T D Evans$^5$, J Rademacker$^6$ and M D Sokoloff$^3$ }

\address{$^1$ CERN, CH-1211 Geneva 23, Switzerland}
\address{$^2$ Experimentelle Physik V, Technische Universit\"at Dortmund, Otto-Hahn-Stra{\ss}e 4, 44227 Dortmund, Germany}
\address{$^3$ Physics Department, University of Cincinnati, 2600 Clifton Ave. Cincinnati, OH 45221, USA}
\address{$^4$ Physikalisches Institut, Universit\"at Heidelberg, Im Neuenheimer Feld 226, 69120 Heidelberg, Germany}
\address{$^5$ Department of Physics, University of Oxford, Parks Road, Oxford OX1 3PU, UK}
\address{$^6$ School of Physics, University of Bristol, Tyndall Avenue, Bristol BS8 1TL, UK}

\ead{christoph.hasse@cern.ch, johannes.albrecht@cern.ch, antonio.augusto.alves.junior@cern.ch, p.dargent@cern.ch, timothy.david.evans@cern.ch, jonas.rademacker@bristol.ac.uk and sokoloff@ucmail.uc.edu}

\begin{abstract}
    The GooFit Framework is designed to perform maximum-likelihood fits for arbitrary functions on various parallel back ends, for example a GPU. We present an extension to GooFit which adds the functionality to perform time-dependent amplitude analyses of pseudoscalar mesons decaying into four pseudoscalar final states. Benchmarks of this functionality show a significant performance increase when utilizing a GPU compared to a CPU. Furthermore, this extension is employed to study the sensitivity on the \(\PDzero-\APDzero\) mixing parameters \(x\) and \(y\) in a time-dependent amplitude analysis of the decay \DKpipipi. Studying a sample of \SI{50000}{} events and setting the central values to the world average of \(x=\SI{0.49\pm0.15}{\percent}\) and \(y=\SI{0.61\pm0.08}{\percent}\), the statistical sensitivities of \(x\) and \(y\) are determined to be \(\sigma(x)=\SI{0.019}{\percent}\) and \(\sigma(y)=\SI{0.019}{\percent}\).

\end{abstract}

\section{Introduction}
In physics analyses it is common to fit a theoretical model to observed data to extract parameters of interest. This involves minimizing the differences between a model and data, which is mostly done by performing a minimization of a cost function, for example the likelihood function. However, problems arise because the computations become very expensive as the complexity of the models and number of events increases. The GooFit~\cite{GF_phys,GF_IEEE,GF_Git} framework has been designed to address this issue by allowing such computations to be performed in parallel. It is built upon the Thrust library~\cite{thrust} to be able to run on different parallel architectures, while maintaining a control flow similar to the RooFit package~\cite{RooFit}, which is commonly used in high energy physics to fit theoretical models to data, and which only runs on CPUs. While GooFit has been successfully employed in several analyses, even for complex models such as time-dependent mixing in three-body decays, it did not allow for performing a time-dependent amplitude analyses of four-body decays. This functionality was recently added and will be described in this paper.

\section{Mixing in the decay \DKpipipi }
Mixing or oscillation of neutral mesons is a process during which a particle transitions into its antiparticle or vice versa. This process has been observed in the \PKzero, \PBzero, \PBs and \PDzero systems. The \PDzero system is the only one comprised of up-type quarks.
\begin{figure}[h!]
\centering
\begin{tikzpicture}[thick]
  \node (A)
       {\PDzero};
  \node (B1) [above right=1.3cm and 2.35cm of A]
       {\(\mathcal{A}_f\)};
  \node (T2) [below right=1cm and .1cm of A]
       {Mixing};
  \node (T3) [below right=.9cm and 4.7cm of A]
       {\(\mathcal{\bar A}_f\)};
  \node (B2) [below right=.9cm and 2.5cm of A]
       {\APDzero};
  \node (C)  [right=5cm of A]
       {\(\PKp\pi^+\pi^-\pi^-\)};

\draw [->] (A) to [out=45,in=135] (C);
\draw [->] (A) to [out=-45,in=180] (B2);
\draw [->] (B2) to [out=0,in=-135] (C);
\end{tikzpicture}
\caption{Schematic view of the two possible decay paths for a \PDzero decaying into a \PKplus\Ppiminus\Ppiplus\Ppiminus final state. The top path corresponds to the direct decay, while the bottom path shows the mixing transition of a \PDzero into a \APDzero followed by a decay into the final state.}
\label{fig:dcsVScf}
\end{figure}
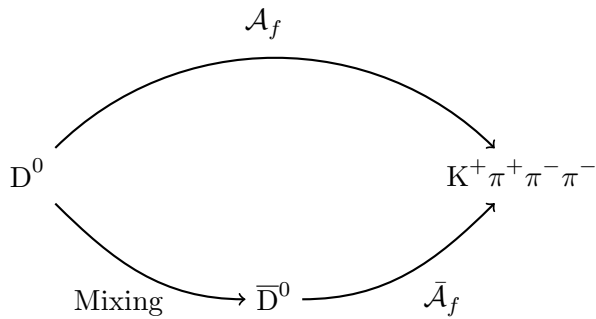
One possible decay to study the phenomenon of mixing in the neutral charm meson system is the decay of \PDzero to \(\PKp\Ppiminus\Ppiplus\Ppiminus\). This decay can proceed via two different decay amplitudes, which are depicted in \cref{fig:dcsVScf}. The top arrow depicts the direct decay subscribed \(\mathcal{A}_f\), while the bottom arrow represents the decay proceeding via mixing into a \APDzero which decays into the final state via an amplitude subscribed \(\mathcal{\bar A}_f\). Due to the mixing of a \PDzero into a \APDzero being time-dependent, the overall decay rate becomes time-dependent. Analysing such time-dependent decay rates allows extraction of mixing properties of the \PDzero system.The expression for the time-dependent decay rate of the \PDzero, assuming no CP violation, can be derived to be~\cite{sozzi},
\begin{align}\label{eq:mix}
    \begin{split}
    	\frac{d\Gamma(\mathcal{A}_f)}{e^{-\Gamma t}\mathcal{N}_f} =& \left(\left|\mathcal{A}_f\right|^2 + \left|\mathcal{\bar A}_f\right|^2\right)\cosh(y\Gamma t) + \left(\left|\mathcal{A}_f\right|^2 - \left|\mathcal{\bar A}_f\right|^2\right)\cos(x\Gamma t) \\ -& 2 \Re\left(\mathcal{A}_f\mathcal{\bar A}_f^*\right)\sinh(y\Gamma t) - 2\Im\left(\mathcal{A}_f\mathcal{\bar A}_f^*\right)\sin(x\Gamma t)\,.
    \end{split}
\end{align}
Most of the complexity of this expression lies within the model used to describe the two amplitudes \(\mathcal{A}_f\) and \(\mathcal{\bar A}_f\)
\section{Structure and implementation of four-body amplitudes}
While \cref{eq:mix} is completely general, the amplitudes that encode the properties of the decay are functions of the position in phase space occupied of the final state of the decay. The amplitude structure of a four-body decay is significantly more complicated than that of three-body decays because their phase space is five dimensional while three-body decays merely occupy a two-dimensional phase space.

Similar to other amplitude models, the implemented functionality assumes that multi-body decays mostly proceed via quasi two-body processes, which include two-body resonances.
\begin{figure}[htpb]
    \centering
    \includegraphics[trim={4.8cm 19cm 2cm 4cm},clip,width=0.7\textwidth]{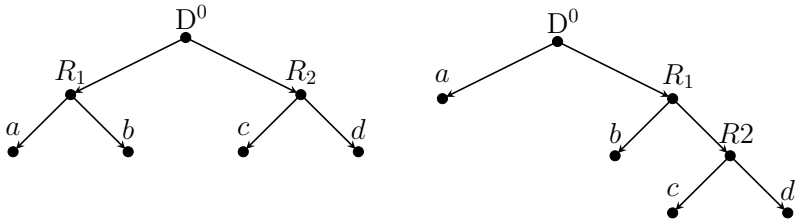}
    \caption{Possible quasi two-body decay topologies of a four-body decay. Left, a \PDzero meson decays into two resonances \(R_1\) and \(R_2\) , which decay into two particles each. Right, a \PDzero meson decays into a particle \(a\) and a resonance \(R_1\) , which proceeds to decay into a resonance \(R_2\) and a final state particle. \(R_2\) then decays into the remaining two final state particles. }
    \label{fig:topo}
\end{figure}
This leads to two possible decay chain topologies depicted in \cref{fig:topo}, where \(R_1\) and \(R_2\) are intermediate resonances and \(a,b,c\) and \(d\) are the four final decay products, in various configurations. Here, \(R_1\) and \(R_2\) can take the form of multiple kinematically allowed resonance states, resulting in many possible decay chains. A complete amplitude will therefore be modelled by a coherent sum over these decay chains \(\mathcal A_i\) as,
\begin{equation}
    \mathcal A_f=\sum_ic_i\mathcal A_i\,, \qquad c_i,\mathcal A_i \in \mathbb{C}.
\end{equation}
Each decay chain \(\mathcal A_i\) is constructed by the user from classes representing form factors, spin factors, resonance lineshapes, and possibly, in the case of two identical final state particles, Bose-symmetrization. After successfully constructing all necessary decay chains the user constructs two amplitude class instances representing \(\mathcal A_f\) and \(\mathcal{\bar A}_f\), which each hold the necessary decay chains to fit the theoretical model. The model creation is finalized by creating an instance of the time-dependent amplitude model class and passing the two amplitudes just created by the user. Upon creation the time-dependent model class automatically checks for recurring form factors, spin factors, and lineshapes in all decay chains. In case of multiple occurrences, these instances are substituted by a link to a single instance, thus removing redundant calculations. The proceeding steps of the internal model building process are explained in detail in~\cite{GF_phys,GF_IEEE}.
\subsection{Normalization and event generation}
During the fitting procedure the complete expression in \cref{eq:mix} must be normalized accurately. As it is not feasible to find an analytic expression for such a complex function, the normalization is computed numerically. In our study, this requires evaluating the function at several million phase space points. To achieve a sufficiently fast generation of phase space events, we integrated the MCBooster library~\cite{MCB_Git,MCB}, which allows very fast generation of phase space events on the GPU. This also enables the generation of pseudo-events, which are uniformly distributed phase space events weighted by the previously created amplitude model.

\subsection{Validation}
As this work implemented various new building blocks to model four-body decay amplitudes in GooFit it was important to validate the correctness of each of these new components. A cross check of the implementation was performed by comparing the newly implemented functionality of GooFit to the software package MINT3~\cite{M3}. MINT3 is based upon the MINT (Minuit Interface) package~\cite{Mint}, which is used to perform time-integrated amplitude analyses of three- and four-body decays. Additionally, it supports the generation of pseudo-events. We generate \(500,000\) pseudo-events for a specific amplitude model, which includes all newly implemented building blocks, and compare the resulting event samples. This comparison is performed by studying the phase space projections of the samples given the five variables \(m_{12}, m_{34}, \cos_{12}, \cos_{34}\) and \(\phi\), where the subscript \(_{12}\) refers to the \Ppiplus\Ppiminus pair and \(_{34}\) to the \PKp\Ppiminus pair.
\begin{figure}[htbp]
	\centering
	\includegraphics[width=0.4\textwidth]{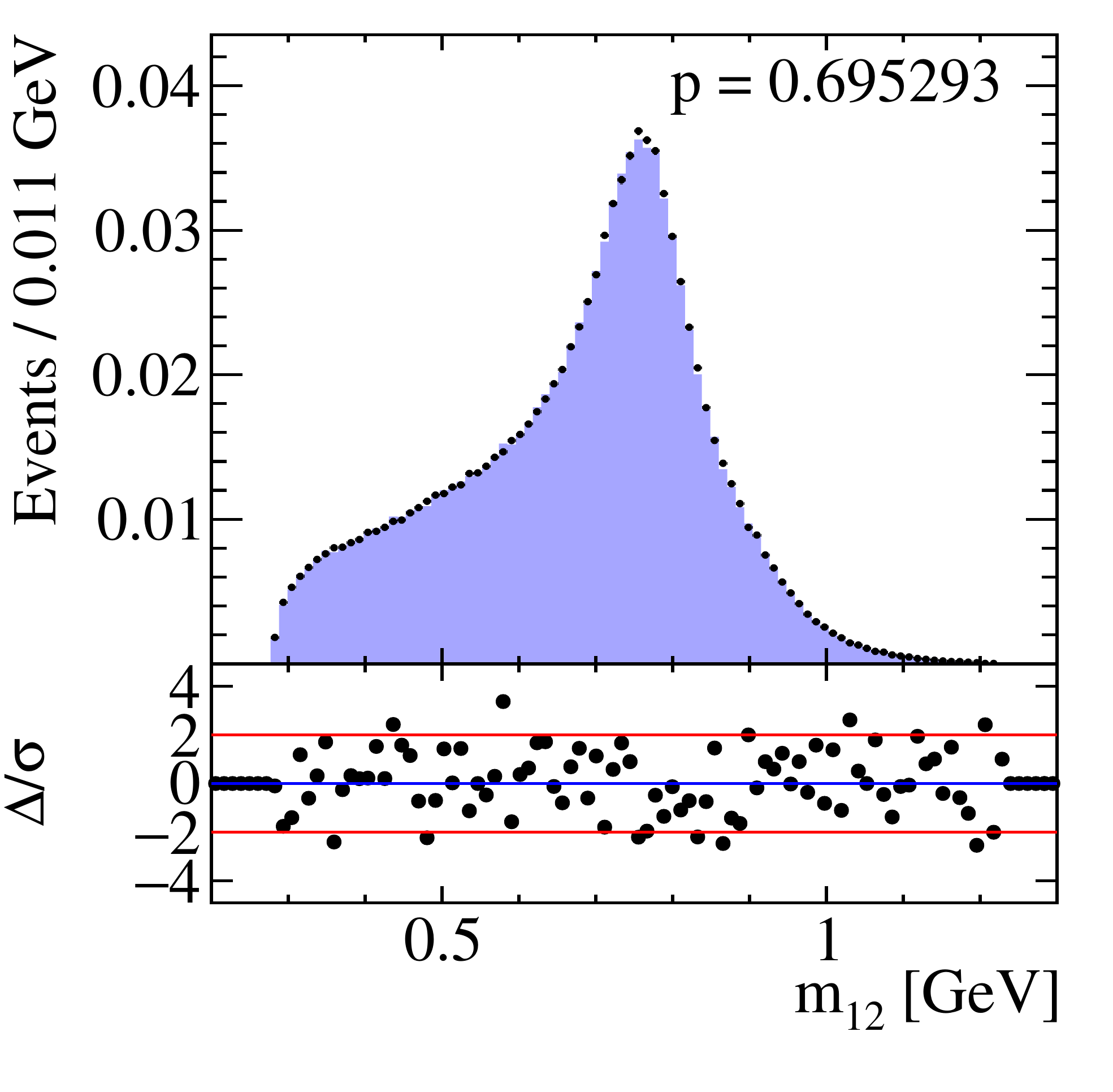}%
	\includegraphics[width=0.4\textwidth]{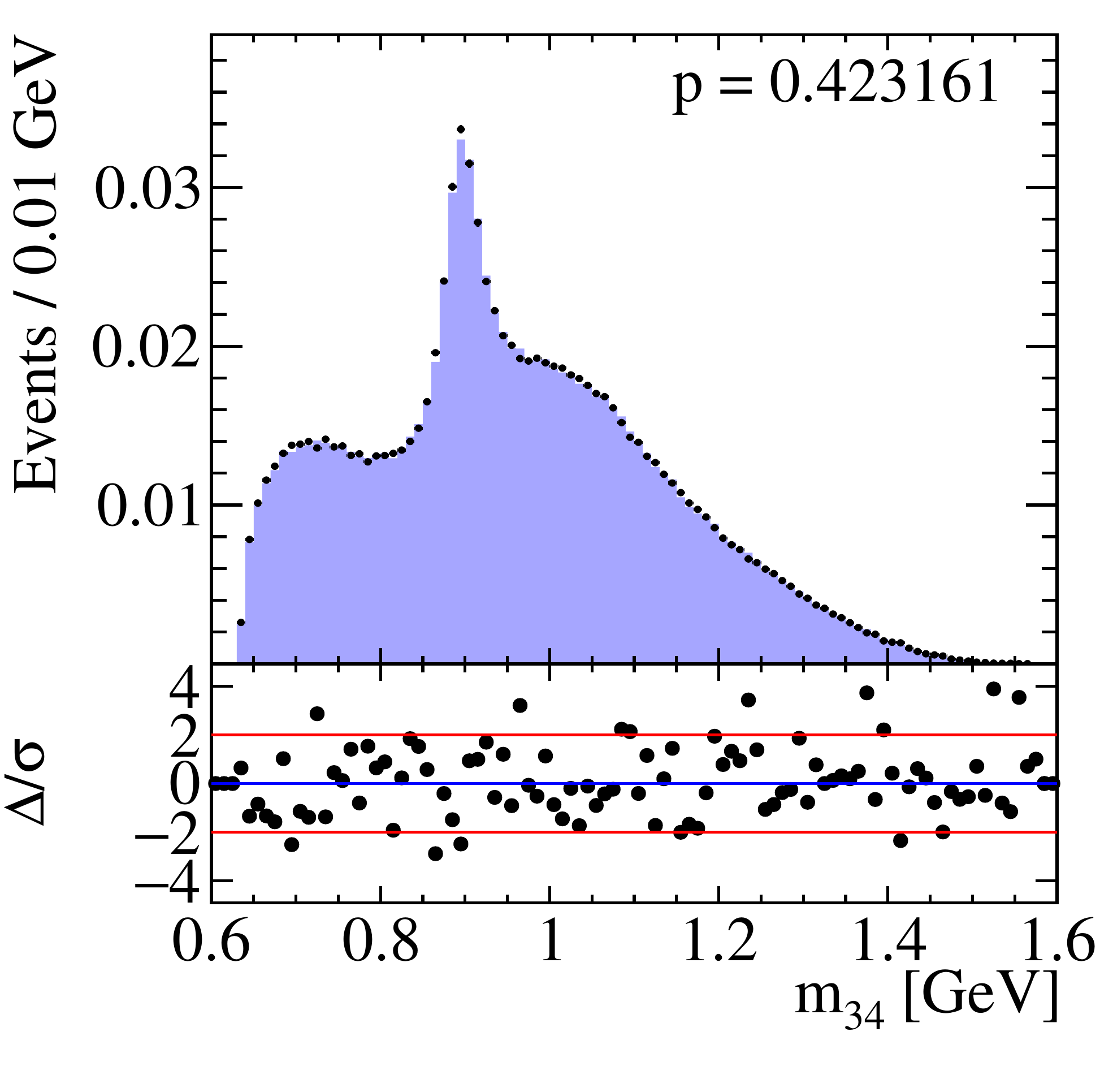}
	\includegraphics[width=0.4\textwidth]{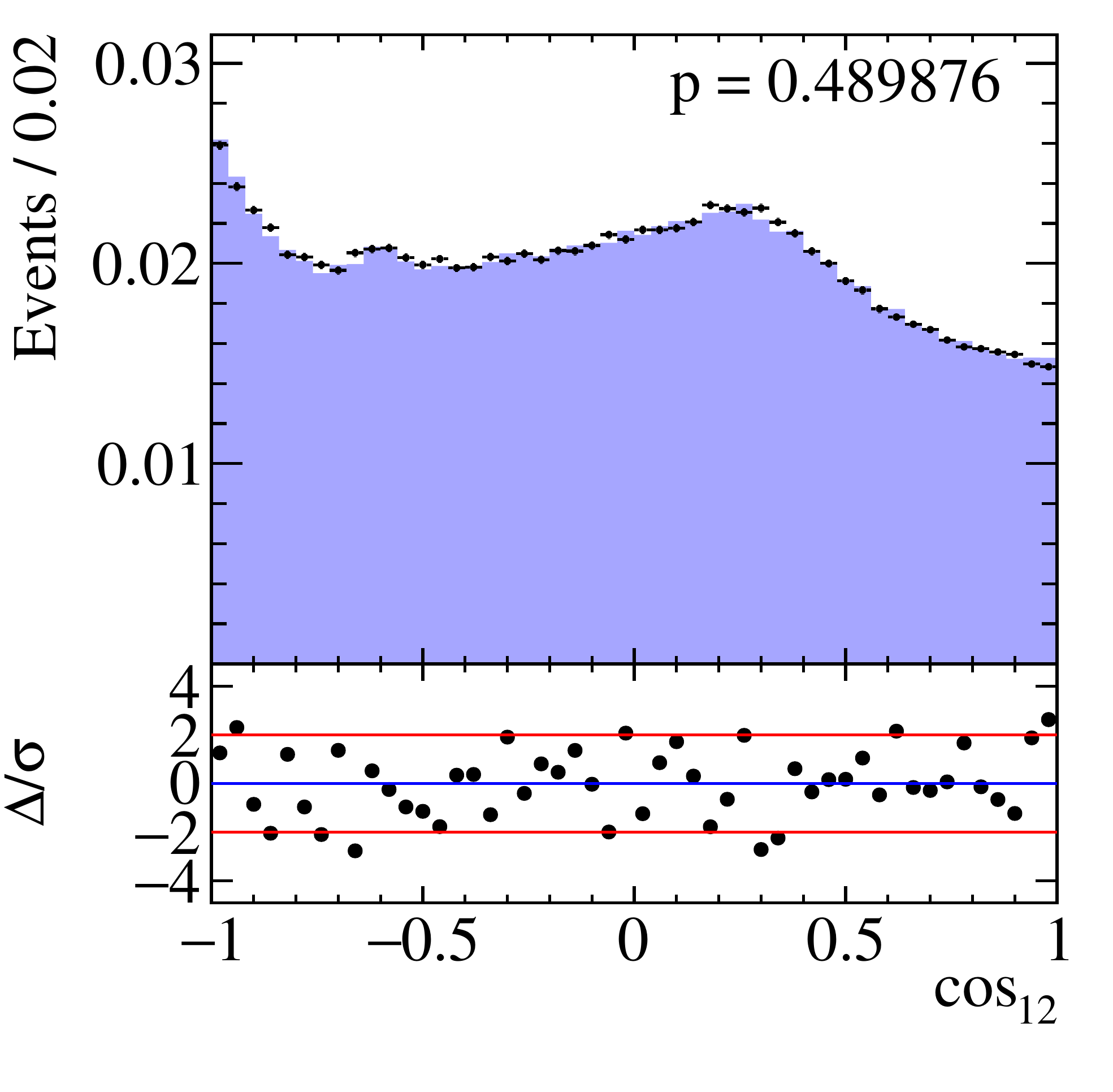}%
	\includegraphics[width=0.4\textwidth]{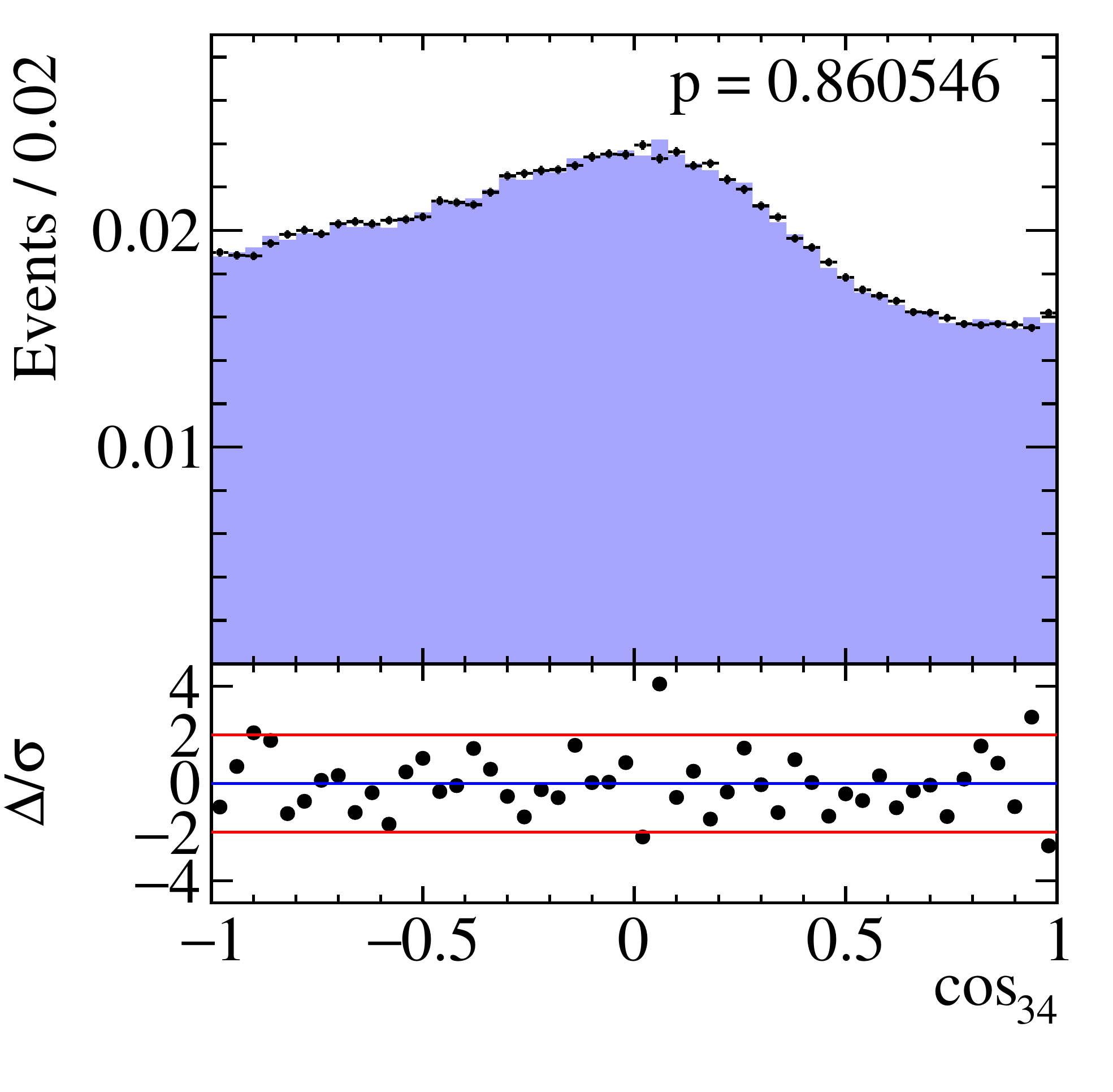}
	\includegraphics[width=0.4\textwidth]{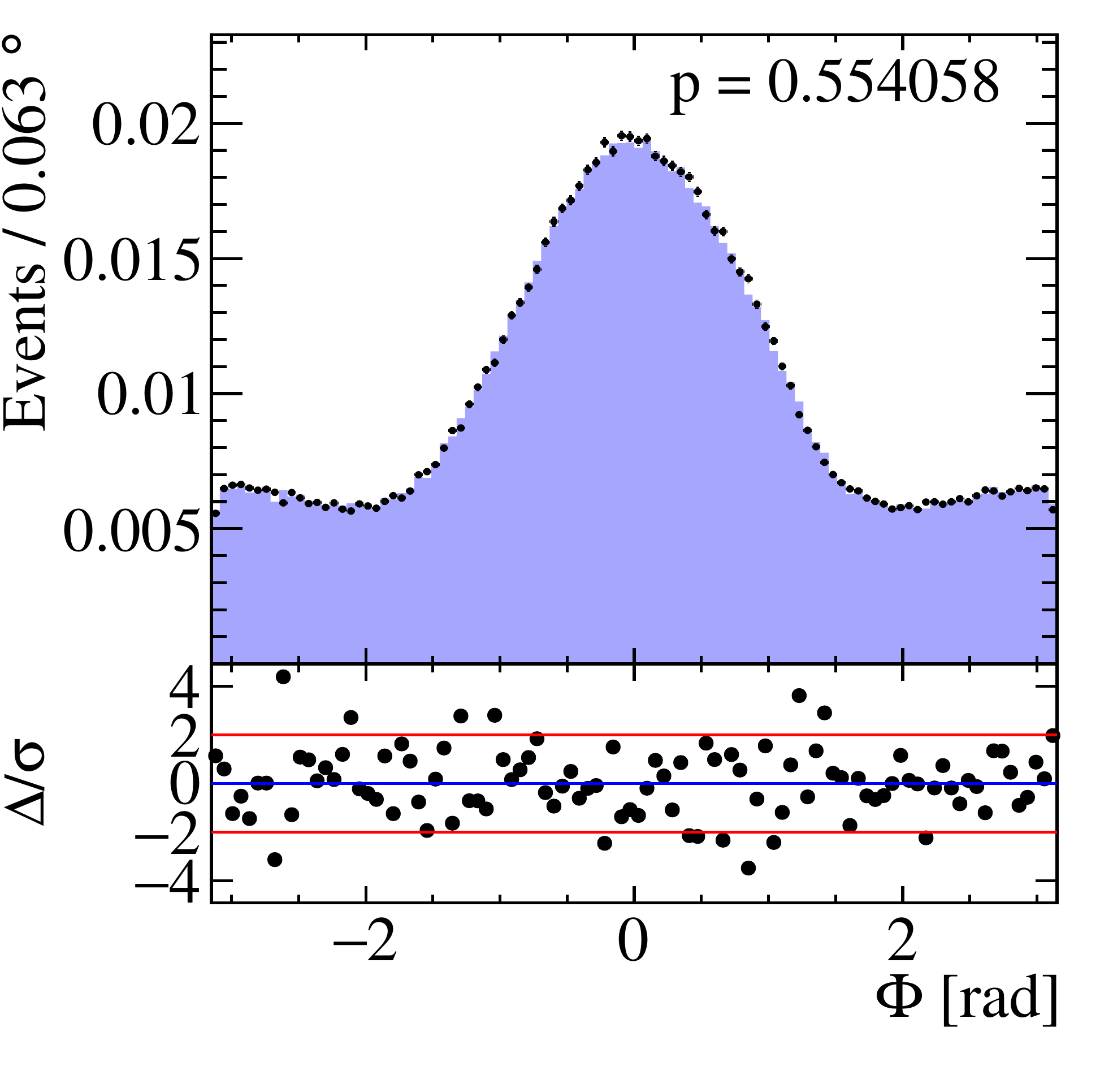}%
	\caption{Comparison between the generated pseudo events from the MINT3 (dots) and GooFit (solid) frameworks. Shown are the distributions of the five variables used to parametrize the phase space. Additionally, the normalized pull distributions and p-value are shown. The pulls should follow a normal distribution with a mean of zero (blue line) and a standard deviation of one. The red lines mark the 2\(\sigma\) region. }
	\label{fig:validation}
\end{figure}
As shown in \cref{fig:validation}, there are no significant differences observed and the pull distribution as well as the p-value indicate that both samples are drawn from the same distribution.

\section{Statistical sensitivity to the charm mixing parameters \(x\) and \(y\)}
The novel functionality of GooFit has successfully been used to determine the statistical sensitivity on the charm mixing parameters \(x\) and \(y\) in a time-dependent amplitude analysis of the decay \DKpipipi. This study did not account for resolution effects, background in the data, and did not allow the model to float. Therefore, the real sensitivity will be worse than shown in \cref{tbl:sens}. Nevertheless, this study proves the capabilities of the newly implemented extension in GooFit to be fully functional.
\begin{table}[htbp]
\centering
    \caption{Summary of the obtained statistical sensitivities of \(x\) and \(y\) in the case of \(x=\SI{0.49}{\percent}\) and \(y=\SI{0.61}{\percent}\)~\cite{hfag}.}
\begin{tabular}{lll}
\toprule
    Events &Sensitivity of \(x\) \([\SI{}{\percent}]\)   & Sensitivity of \(y\) \([\SI{}{\percent}]\) \\
\midrule
20000 & \SI{0.030}{}&\SI{0.031}{} \\
50000 & \SI{0.019}{}&\SI{0.019}{}  \\
70000 &\SI{0.016}{}&\SI{0.017}{}\\
\bottomrule
\end{tabular}
\label{tbl:sens}
\end{table}
\section{Performance comparison between CPU and GPU}
Lastly, we present a performance comparison of the newly implemented functionality, between the CPU and GPU. Two different test cases are used to study the performance. The first one targets the generation speed of pseudo-events according to a time-dependent amplitude-model. This generation is repeated for three different sample sizes to study the scaling behavior. Secondly, the performance of the fitting procedure is studied, where the scaling behavior is studied by increasing the number of used events in the normalization while leaving the sample size one fits to constant.

These tests are repeated on three different platforms: a server with two Intel Xeon E5-2680 v3 CPUs, each with 12 physical cores that can run two concurrent threads, a NVIDIA K40 GPU and a mid-range mobile gaming GPU NVIDIA GeForce GT 525M. The results are obtained by an average over 5 runs, and listed in \cref{tab:perf1,tab:perf2}. They show a significant speedup when utilizing the K40 and even the outdated mid-range mobile graphics card was able to perform surprisingly well compared to the other two platforms, but due to insufficient memory it was not able to complete all tests.
\begin{table}[!h]
\caption{Pseudo-event generation according to a time-dependent model using a Monte-Carlo accept/reject method.}
        \centering
        \begin{tabular}{lllll}
        \toprule
        \multirow{3}{*}{Events} & \multicolumn{2}{c}{2 \(\times\) Intel Xeon} & \multicolumn{2}{c}{NVIDIA} \\
                     & \multicolumn{2}{c}{E5-2680 v3 2.50GHz} & GT 525M & K40 \\
                     & 24 Cores & 48 Cores & 96 Cores &         2880 Cores \\
        \midrule
        \SI{20000}{} &\SI{179.7}{\second} &\SI{156.7}{\second} & \SI{195.8}{\second} &\SI{25.0}{\second} \\
        \SI{50000}{} &\SI{451.9}{\second} &\SI{378.7}{\second} & \SI{484.6}{\second} &\SI{58.8}{\second} \\
        \SI{70000}{} &\SI{598.0}{\second} &\SI{524.0}{\second} & \SI{677.4}{\second} &\SI{79.0}{\second}\\ \bottomrule
        \end{tabular}
        \label{tab:perf1}
\end{table}
\begin{table}[!h]
    \lineup
    \caption{Fit to \SI{100000}{} generated pseudo-events, with varying number of points used to calculate the normalization. Fixed model, floating \(x\) and \(y\).}
        \centering
        \begin{tabular}{lllll}
        \toprule
        \multirow{3}{*}{Points} & \multicolumn{2}{c}{2 \(\times\) Intel Xeon} & \multicolumn{2}{c}{NVIDIA} \\
                     & \multicolumn{2}{c}{E5-2680 v3 2.50GHz} & GT 525M & K40  \\
                     & 24 Core & 48 Cores & 96 Cores & 2880 Cores \\
        \midrule
        \m\SI{750000}{} & \SI{4.2}{\second}  & \SI{3.3}{\second}  & \SI{8.2}{\second} & \SI{0.6}{\second} \\
        \SI{1500000}{}  & \SI{7.7}{\second}  & \SI{6.5}{\second}  & -                 & \SI{1.0}{\second} \\
        \SI{3000000}{}  & \SI{14.8}{\second} & \SI{12.4}{\second} & -                 & \SI{1.7}{\second} \\
        \SI{6000000}{}  & \SI{30.0}{\second} & \SI{22.5}{\second} & -                 & \SI{3.2}{\second} \\ \bottomrule
        \end{tabular}
        \label{tab:perf2}
\end{table}

While the non-linear scaling from 24 to 48 cores was expected as one only increases the logical number of cores by running two threads per core, the expected performance gain from the K40 compared to the GT 525M was less than a priori expected. Using the available NVIDIA profiler, we are able to determine that the source of the throttled performance on the K40 is due to memory latency. We hope to reduce this in the future by reducing the used memory as well adapting the current memory layout to make memory transfers more efficient.
\section{Summary}
In conclusion, we have presented a novel extension to the GooFit framework which allows for performing a time-dependent amplitude analysis of a pseudoscalar meson decaying into four pseudo-scalar final states. Additionally, this extension allows the user to generate pseudo-events according to a previously defined time-dependent amplitude model. This functionality was successfully validated by comparing the results to an existing software package and furthermore used to study the sensitivity to the charm mixing parameters in the decay \DKpipipi. Lastly, it is shown that there is a significant speedup gained by utilizing the GPU, while an even bigger performance gain is forseen once the memory layout in GooFit has been adapted to minimize memory latency on high performance GPUs like the K40.

The GooFit package can be found on GitHub at \url{https://github.com/GooFit}
\ack
I would like to thank  the authors and maintainers of the MINT and MINT3 framework, P. d'Argent, T.D. Evans and J. Rademacker, as their work and support has been most helpful in implementing the presented extension to GooFit.

Work sponsored by the Wolfgang Gentner Programme of the Federal Ministry of Education and Research.

The development of this extension has been in part supported by the National Science Foundation under grant number PHY-1414736.

NVidia provided K40 GPUs for our use through its University Partnership program.
\bibliography{references}

\providecommand{\newblock}{}
\begin{thebibliography}{10}
\expandafter\ifx\csname url\endcsname\relax
  \def\url#1{{\tt #1}}\fi
\expandafter\ifx\csname urlprefix\endcsname\relax\def\urlprefix{URL }\fi
\providecommand{\eprint}[2][]{\url{#2}}

\bibitem{GF_phys}
Andreassen R, Meadows B~T, de~Silva M, Sokoloff M~D and Tomko K 2014 {\em
  Journal of Physics: Conference Series\/} {\bf 513} 052003

\bibitem{GF_IEEE}
Andreassen R~E, de~Silva W~M, Meadows B~T, Sokoloff M~D and Tomko K~A 2014 {\em
  IEEE Access\/} {\bf 2} 160--176

\bibitem{GF_Git}
{The GooFit package} \url{https://github.com/GooFit/GooFit}

\bibitem{thrust}
{The Thrust library} \url{https://thrust.github.io/}

\bibitem{RooFit}
Verkerke W and Kirkby D~P 2003 {\em eConf\/} {\bf C0303241} (\textit{Preprint}
  \eprint{physics/0306116})

\bibitem{sozzi}
Sozzi M~S 2008 {\em {Discrete symmetries and CP violation: From experiment to
  theory}\/}

\bibitem{MCB_Git}
{The MCBooster library} \url{https://github.com/MultithreadCorner/MCBooster}

\bibitem{MCB}
Alves~Jr A~A {\em et~al.\/} 2016 {MCBooster: a library for fast Monte Carlo
  generation of phase-space decays in massively parallel platforms.}
  \url{http://indico.cern.ch/event/505613/contributions/2230884/}

\bibitem{M3}
{LHCb Collaboration} Analysis in progress

\bibitem{Mint}
Artuso M {\em et~al.\/} (CLEO) 2012 {\em Phys. Rev.\/} {\bf D85} 122002
  (\textit{Preprint} \eprint{1201.5716})

\bibitem{hfag}
Amhis Y {\em et~al.\/} (Heavy Flavor Averaging Group (HFAG)) 2014  Updated
  results and plots available at \url{http://www.slac.stanford.edu/xorg/hfag/}.
  (\textit{Preprint} \eprint{1412.7515})

\end{thebibliography}

\end{document}